\documentclass[]{aastex631}

\usepackage{amsmath}

\received{September 14, 2021}
\revised{November 11, 2021}
\accepted{November 12, 2021}

\submitjournal{The Astronomical Journal}

\begin{document}

\title{An algorithm to calibrate and correct the response to unpolarized radiation of the X-ray polarimeter on board IXPE}

\correspondingauthor{John Rankin}
\email{john.rankin@inaf.it}

\author[0000-0002-9774-0560]{John Rankin}
\affiliation{Istituto di Astrofisica e Planetologia Spaziali di Roma, Via Fosso del Cavaliere 100, I-00133 Roma, Italy}
\affiliation{Università di Roma Sapienza, Dipartimento di Fisica, Piazzale Aldo Moro 2, 00185 Roma, Italy}
\affiliation{Università di Roma Tor Vergata, Dipartimento di Fisica, Via della Ricerca Scientifica, 1, 00133 Roma, Italy}

\author[0000-0003-3331-3794]{Fabio Muleri}
\affiliation{Istituto di Astrofisica e Planetologia Spaziali di Roma, Via Fosso del Cavaliere 100, I-00133 Roma, Italy}

\author{Allyn F. Tennant}
\affiliation{NASA Marshall Space Flight Center, Huntsville, AL 35812, USA}

\author[0000-0002-4576-9337]{Matteo Bachetti}
\affiliation{Istituto di Astrofisica e Planetologia Spaziali di Roma, Via Fosso del Cavaliere 100, I-00133 Roma, Italy}

\author[0000-0003-4925-8523]{Enrico Costa}
\affiliation{Istituto di Astrofisica e Planetologia Spaziali di Roma, Via Fosso del Cavaliere 100, I-00133 Roma, Italy}

\author[0000-0003-0331-3259]{Alessandro Di Marco}
\affiliation{Istituto di Astrofisica e Planetologia Spaziali di Roma, Via Fosso del Cavaliere 100, I-00133 Roma, Italy}

\author[0000-0003-1533-0283]{Sergio Fabiani}
\affiliation{Istituto di Astrofisica e Planetologia Spaziali di Roma, Via Fosso del Cavaliere 100, I-00133 Roma, Italy}

\author[0000-0001-8916-4156]{Fabio La Monaca}
\affiliation{Istituto di Astrofisica e Planetologia Spaziali di Roma, Via Fosso del Cavaliere 100, I-00133 Roma, Italy}

\author[0000-0002-7781-4104]{Paolo Soffitta}
\affiliation{Istituto di Astrofisica e Planetologia Spaziali di Roma, Via Fosso del Cavaliere 100, I-00133 Roma, Italy}

\author[0000-0001-9194-9534]{Antonino Tobia}
\affiliation{Istituto di Astrofisica e Planetologia Spaziali di Roma, Via Fosso del Cavaliere 100, I-00133 Roma, Italy}

\author[0000-0002-3180-6002]{Alessio Trois}
\affiliation{INAF/Osservatorio Astronomico di Cagliari, Via della Scienza 5, I-09047 Selargius (CA), Italy}

\author[0000-0002-0105-5826]{Fei Xie}
\affiliation{Istituto di Astrofisica e Planetologia Spaziali di Roma, Via Fosso del Cavaliere 100, I-00133 Roma, Italy}

\author[0000-0002-9785-7726]{Luca Baldini}
\affiliation{Università di Pisa, Dipartimento di Fisica Enrico Fermi, Largo B. Pontecorvo 3, I-56127 Pisa, Italy}
\affiliation{Istituto Nazionale di Fisica Nucleare, Sezione di Pisa, Largo B. Pontecorvo 3, I-56127 Pisa, Italy}

\author{Niccolò Di Lalla}
\affiliation{W.W. Hansen Experimental Physics Laboratory, Kavli Institute for Particle Astrophysics and Cosmology, Department of Physics and SLAC National AcceleratorLaboratory, Stanford University, Stanford, CA 94305, USA}

\author[0000-0002-0998-4953]{Alberto Manfreda}
\affiliation{Istituto Nazionale di Fisica Nucleare, Sezione di Pisa, Largo B. Pontecorvo 3, I-56127 Pisa, Italy}

\author[0000-0002-1868-8056]{Stephen L. O'Dell}
\affiliation{NASA Marshall Space Flight Center, Huntsville, AL 35812, USA}

\author[0000-0003-3613-4409]{Matteo Perri}
\affiliation{INAF/Osservatorio Astronomico di Roma, Via Frascati 33, I-00040, Monte Porzio Catone (RM)}

\author[0000-0002-2734-7835]{Simonetta Puccetti}
\affiliation{Agenzia Spaziale Italiana, Via del Politecnico snc, I-00133 Roma, Italy}

\author{Brian D. Ramsey}
\affiliation{NASA Marshall Space Flight Center, Huntsville, AL 35812, USA}

\author[0000-0001-5676-6214]{Carmelo Sgrò}
\affiliation{Istituto Nazionale di Fisica Nucleare, Sezione di Pisa, Largo B. Pontecorvo 3, I-56127 Pisa, Italy}

\author[0000-0002-5270-4240]{Martin C. Weisskopf}
\affiliation{NASA Marshall Space Flight Center, Huntsville, AL 35812, USA}

\begin{abstract}

The Gas Pixel Detector is an X-ray polarimeter to fly on-board IXPE and other missions. 
To correctly measure the source polarization, the response of IXPE's GPDs to unpolarized 
radiation has to be calibrated and corrected. In this paper we describe the way such response
is measured with laboratory sources and the algorithm to 
apply such correction to the observations of celestial sources. 
The latter allows to correct the response to polarization of single photons, 
therefore allowing great flexibility in all the subsequent analysis. 
Our correction approach is tested against both monochromatic
and non-monochromatic laboratory sources and with simulations, finding that 
it correctly retrieves the polarization up to the statistical limits of the planned IXPE observations.

\end{abstract}

\keywords{Polarimeters(1277) --- X-ray telescopes(1825) --- X-ray observatories(1819) --- X-ray detectors(1815)}

\section{Introduction}

Astronomical X-ray polarimetry has up until now seen significative
detections only of the Crab Nebula \citep{1978ApJ...220L.117W, 2020NatAs...4..511F}, 
but this unexplored window will
soon be reopened thanks to the IXPE mission \citep{2016SPIE.9905E..17W, Soffitta_2021},
with on board the polarization sensitive Gas Pixel Detector (GPD) 
\citep{2001Natur.411..662C,2006NIMPA.566..552B,2007NIMPA.579..853B}.
This device has already flown on-board the PolarLight cubesat mission 
\citep{2019ExA....47..225F}, providing new results on the Crab Nebula, 
and it will also fly on future missions (e.g., eXTP \citep{2019SCPMA..6229502Z}).

Expected polarization from X-ray astronomical sources is higher than at 
longer wavelengths, e.g., optical or infrared, but still a few \% of the 
source signal. At this level IXPE's GPDs, and often other real X-ray 
polarimeters, show systematic effects that mimic the signal generated 
by a genuine source polarization even for truly unpolarized radiation \citep{baldini2021}.
Due to the characteristics of this effect, and to facilitate its correction, 
part of it will be compensated by the fact that IXPE's observations will 
be dithered (that is the pointing direction of the telescope will oscillate, 
during the observations, distributing source photons over a relatively 
large region, nearly uniformly illuminated and centered on the field of view).
 The remainder of these systematic effects need to be calibrated 
\citep{spurious_future} and removed before being able to achieve 
the statistical limit of the polarization measurement.

In this paper we describe the algorithm to calibrate and correct the response of 
IXPE's GPDs to unpolarized radiation so to remove the instrumental spurious 
signal, named spurious modulation.
First, we describe how the response of the detector to unpolarized
radiation is measured. This method gives, as a byproduct, the corrected
polarization of the laboratory sources.

We then present the algorithm that will be used for the correction
of celestial observations. It is able to remove the systematic effect
from individual photon events, therefore rendering the subsequent
analysis flexible. This method comprises two parts: the creation of
a calibration database containing spatial and spectral information
on the systematic effect, and the latter subtraction from each single
photon detected by using the spatial information and interpolating
the spectral information. 

We compare, for monochromatic sources, the polarization obtained by using
this correction algorithm with that obtained as a byproduct of the
unpolarized response measurement. We
also study the application of the correction method to 
non-monochromatic sources.  The algorithm is further tested 
using toy simulations to verify that statistical uncertainties are propagated correctly.

This paper is structured as follows. In Sec. \ref{sec:GPD} we
describe how the GPD measures X-ray polarization. The method used
to measure the response to unpolarized radiation is presented in Sec.
\ref{sec:unpol_response}, while the photon by photon correction algorithm
is described in Sec. \ref{sec:ph_by_ph}. The testing of this algorithm
applied to laboratory X-ray sources and to simulations is reported in 
Sec. \ref{sec:calibration_testing}. Finally, in Sec. \ref{sec:conclusion}, 
conclusions are drawn.

\section{The Gas Pixel Detector\label{sec:GPD}}

The Gas Pixel Detector (GPD) is
a polarization sensitive X-ray detector able to perform spatially, timing, 
and spectrally-resolved polarization measurements. This detector 
\citep{baldini2021} is the core of the IXPE instrument \citep{Soffitta_2021},
and is scheduled to fly on board the IXPE mission. The measurements in
this article were acquired at INAF-IAPS during the ground calibration of 
the Detector Units \citep{spurious_future, ixpe_cal},
with INAF-IAPS's calibration equipment \citep{MULERI2022102658}.
The IXPE instrument consists of three Detector Units (each containing
a GPD) with three corresponding X-ray optics. The data used in this paper 
was taken from detector unit number
2, and is representative of the results obtained with the other detector
units.

\begin{figure}
\begin{centering}
\includegraphics[width=0.7\linewidth]{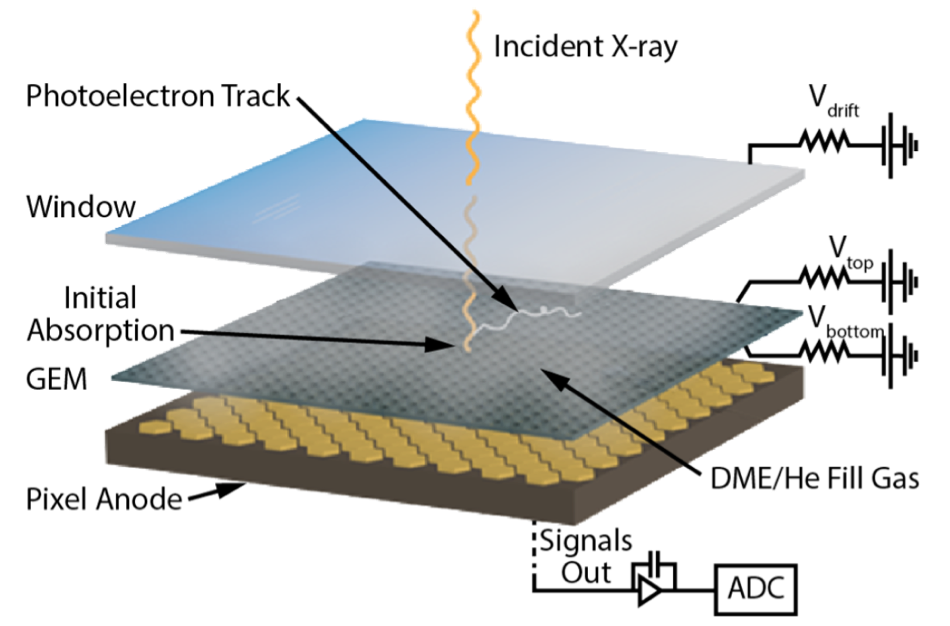}
\par\end{centering}
\caption{Schematic of the GPD. See the text for details. Image credit: \citet{2016SPIE.9905E..17W}
\label{fig:GPD_schematic}}
\end{figure}

A schematic view of the GPD is shown in Fig. \ref{fig:GPD_schematic}.
The functioning is the following. An incident X-ray enters through
the beryllium window and is absorbed in the gas cell filled with DME.
The absorption of the photon causes the production of a photoelectron,
which propagates in the gas producing an ionization track. Such a
charge is collected with a drift field, multiplied by a Gas Electron
Multiplier and eventually collected by a custom ASIC specifically
developed for these detectors.

\subsection{Polarization measurement}

The direction of emission of photoelectrons is statistically more
probable parallel to the position angle of the polarization. Therefore the response of
the instrument is essentially the distribution of the (azimuthal) directions
of the photoelectrons, named modulation curve. In case the
incident radiation is unpolarized, the distribution is expected to be
flat, except for statistical fluctuations due to the finite number
of acquired photons (Fig. \ref{fig:modulation_curves} upper right).
In case the incident radiation is polarized, the distribution of photoelectrons
is expected to be modulated as a $\cos^{2}$ (Fig. \ref{fig:modulation_curves}
left).

The ``classical'' approach to derive the polarization degree and angle
is to fit the distribution as $N\left(\phi\right)=A+B\cos^{2}\left(\varphi-\varphi_{0}\right)$
and derive the so called modulation from the parameters of the fit
as $m=\frac{\max(N)-\min(N)}{\max(N)+\min(N)}=\frac{B}{2A+B}$. The
polarization degree is then given by 
\begin{equation}
p=\frac{m}{\mu\left(E\right)}\label{eq:p}
\end{equation}
where $\mu$ is the modulation factor, equal to the modulation
when the incident radiation is 100\% polarized and dependent on the
energy $E$. The polarization position angle $\varphi_{0}$ coincides with
the peak of the modulation curve.

However this approach has two shortcomings with respect to the goal
of this paper. First, because the polarization degree and phase are
both not statistically independent and not additive, the subtraction of systematic effects, such as those treated
in this paper, requires the use of Stokes parameters, which are statistically independent. Second, we are
interested in applying the correction as early as possible in the processing of the data, even
at the level of the single event before polarization determination,
to leave greater flexibility in the subsequent analysis. For these
reasons, we deemed more appropriate not to follow the classical approach,
but to start from the Stokes parameters of the single events.

The approach used in this paper to compute polarization is derived from \citet{2015APh....68...45K}, 
with the minor difference of a factor of 2 in the definition of the Stokes parameters. 
For each photon with photoelectric position angle $\phi_{i}$
the $q_{i}$ and $u_{i}$ Stokes parameters are computed
\begin{equation}
q_{i}=2\cos\left(2\phi_{i}\right)\label{eq:q_i}
\end{equation}
\begin{equation}
u_{i}=2\sin\left(2\phi_{i}\right)\label{eq:u_i}
\end{equation}

Each event has a known spatial position and spectral energy. The events
are therefore selected inside the range desired, and for the $N$
selected events the normalized $q$ and $u$ Stokes parameters are
computed
\begin{equation}
q=\frac{\sum_{i}q_{i}}{N}\label{eq:q}
\end{equation}
\begin{equation}
u=\frac{\sum_{i}u_{i}}{N}\label{eq:u}
\end{equation}
with uncertainty given by the standard deviation
\begin{equation}
 \sigma_q =\sqrt{\frac {2-q^2} {N-1} } \label{eq:q_err}
 \end{equation} 
\begin{equation}
\sigma_u = \sqrt{\frac { 2-u^2} {N-1}} \label{eq:u_err}
\end{equation} 

From these the modulation can be obtained
\begin{equation}
m=\sqrt{q^{2}+u^{2}}\label{eq:m}
\end{equation}

The polarization degree is then given by Eq. \ref{eq:p}, while
the polarization position angle is given by
\begin{equation}
\varphi=\frac{1}{2}\tan^{-1}\left(\frac{u}{q}\right)\label{eq:psi}
\end{equation}

While the Stokes parameters ($q$, $u$) have statistical uncertainties 
($\sigma_q$, $\sigma_u$) that are gaussian distributed and uncorrelated, 
the modulation and position angle ($m$, $\varphi$) do not. 
However, for measurements of high statistical significance, the uncertainties 
($\sigma_m$, $\sigma_{phi}$) become approximately gaussian distributed and uncorrelated, with
	\begin{equation}
\sigma_{m} \approx \sqrt{\frac{2-m^{2}}{N-1}}\label{eq:m_err}
\end{equation}
and
	\begin{equation}
\sigma_{\varphi} \approx \frac{1}{m\sqrt{2\left(N-1\right)}}\label{eq:psi_err}
\end{equation}
for $m/\sigma_m \gg 1$.

These uncertainties are valid without the application of the systematic 
corrections described in this paper, but represent nonetheless a 
good approximation for celestial observations of high statistical significance. 
The complete expressions are described below in Sec. \ref{subsec:uncertainty}, 
where the corrected Stokes parameters of any systematic effects are simply taken into account.

\subsection{Systematic effects}

For unpolarized radiation, an ideal polarimeter would measure only a very small amplitude of
modulation due to the Poisson distribution of photoelectrons and decreasing
as the number of counts increases; however this is not found to be
the case with the GPD, in which a systematic signal, named spurious modulation, is detectable.

\begin{figure}
\begin{centering}
\includegraphics[width=0.7\linewidth]{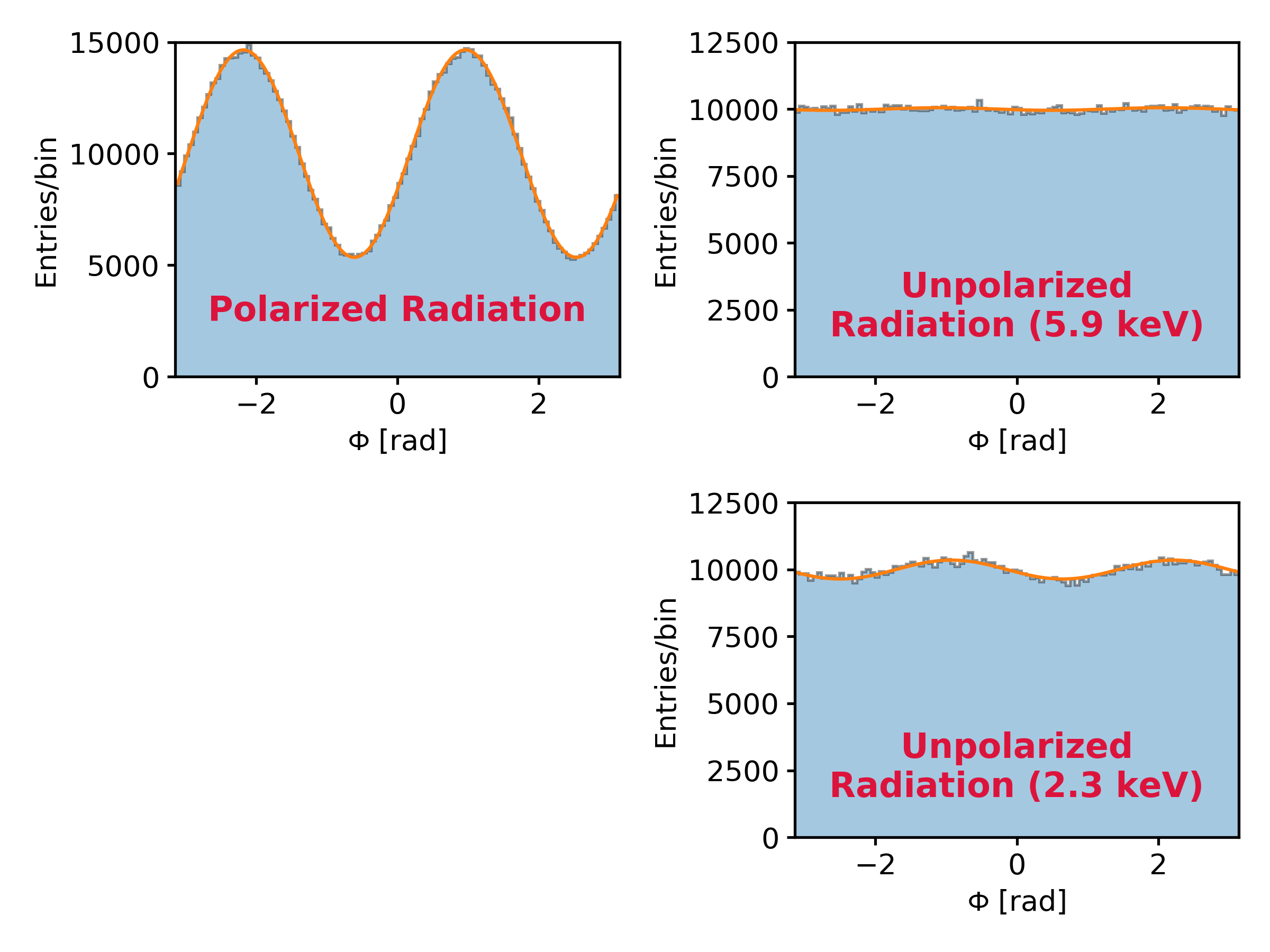}
\par\end{centering}
\caption{Modulation curves for polarized (left), unpolarized at 5.9~keV (upper
right) and unpolarized at 2.3~keV (bottom right) radiation. For polarized
radiation a modulation is seen, which is absent for unpolarized radiation
at high energies, but reappears at low energies as spurious modulation.
\label{fig:modulation_curves}}
\end{figure}

Fig. \ref{fig:modulation_curves} shows the impact of spurious modulation: For polarized
radiation (left) a clear modulation is seen, which is absent for unpolarized
radiation at high energies (upper right), for which a curve very close
to flat is seen. However, for unpolarized radiation at lower energies
(bottom right), a modulated component appears as spurious
modulation.

It should be noted that spurious modulation behaves essentially as an additional $\cos^{2}$ contribution,
that is, it has the same frequency as the modulation caused by a genuine source polarization.
As a consequence, the sum of genuine and spurious modulation will still be a $\cos^{2}$. For this
reason, and depending on the phase, spurious modulation might appear ``hidden'' from the polarized
modulation curve, but is actually present, changing the measured modulation
and shifting the observed phase.

As presented by \citet{spurious_future}, spurious modulation is constant within the calibration requirements of 0.3\% over variations in time, temperature and source rate, and as a consequence ground calibration measurements can be used to correct flight data.
The same paper presents the quantitative details on the effect of spurious modulation on IXPE observations, given the actual calibration measurements available for each detector unit of IXPE.

\subsection{Statistical treatment in the presence of systematic effects}

In this section the formalism of \citet{2015APh....68...45K} to compute expected
values and variances is extended to include spurious modulation, evaluating the case 
of a second component in the polarimetric signal.

The probability distribution function for the photoelectron emission angles 
in case of an ideal X-ray polarimeter is given by
\begin{equation}
f\left(\varphi\right)=\frac{1}{2\pi}\left[1+\mu p_0\cos\left(2\left(\varphi-\varphi_0\right)\right)\right].
\end{equation}
Considering the $q_i$ and $u_i$ Stokes parameters values for a single event 
(Eq. \ref{eq:q_i} and Eq. \ref{eq:u_i})
the expected   values for the Stokes parameters $q_0$ and $u_0$ can be estimated by the relations
\begin{eqnarray}
E[q_{0}] & = & \int_{0}^{2\pi}f\left(\varphi\right)2\cos\left(2\varphi\right)\text{d}\varphi = \mu p_0\cos\left(2\varphi_{0}\right)\\
E[u_{0}]  & = & \int_{0}^{2\pi}f\left(\varphi\right)2\sin\left(2\varphi\right)\text{d}\varphi=\mu p_0\sin\left(2\varphi_{0}\right)
\end{eqnarray}
In the same way the variances can be evaluated
\begin{equation} \label{eq:q0_var}
\begin{aligned}
\text{Var}[q_{0}] & = & \int_{0}^{2\pi}f\left(\varphi\right)\left(2\cos\left(2\varphi\right)-E[q_0]\right)^{2}\text{d}\varphi \\
& = & 2-\left(\mu p_0\cos\left(2\varphi_{0}\right)\right)^{2} = 2 - E[q_{0}]^2
\end{aligned}
\end{equation}
\begin{equation}\label{eq:u0_var}
\begin{aligned}
\text{Var}[u_{0}] & = & \int_{0}^{2\pi}f\left(\varphi\right)\left(2\sin\left(2\varphi\right)- E[u_0] \right)^{2}\text{d}\varphi \\
& = & 2-\left(\mu p_0\sin\left(2\varphi_{0}\right)\right)^{2} = 2 -  E[u_{0}]^2
\end{aligned}
\end{equation}

To extend this approach to take into account a second $\cos^2( \varphi)$ component, 
as expected for spurious modulation, we can define a new probability distribution function
\begin{equation}
g\left(\varphi\right)=\epsilon_{sm}\cos\left(2\left(\varphi-\varphi_{sm}\right)\right)
\end{equation}
where $\epsilon_{sm}$ and $\varphi_{sm}$ are respectively the amplitude and 
phase of spurious modulation. The new overall probability distribution function is
\begin{equation}
F\left(\varphi\right)=\frac{1}{2\pi} \left[2\pi f\left(\varphi\right)+g\left(\varphi\right)\right]
\end{equation}

From this distribution the new expected values for $q$ and $u$ are obtained
\begin{equation}
\begin{aligned}
E[q] & = & \int_{0}^{2\pi}F\left(\varphi\right)2\cos\left(2\varphi\right)\text{d}\varphi\\
& = &  E[q_{0}] +\epsilon_{sm}\cos\left(2\varphi_{sm}\right)  = E[q_{0}] +  E[q_{sm}]
\end{aligned}
\end{equation}
\begin{equation}
\begin{aligned}
E[u] &  = & \int_{0}^{2\pi}F\left(\varphi\right)2\sin\left(2\varphi\right)\text{d}\varphi \\
& = & E[u_{0}] +\epsilon_{sm}\sin\left(2\varphi_{sm}\right) =  E[u_{0}] +  E[u_{sm}]
\end{aligned}
\end{equation}
These results show that spurious modulation is an additional summative term 
to the expected value, and therefore can be subtracted
\begin{eqnarray}
 q_{corr} = q_{meas} - q_{sm} \\ 
 u_{corr} = u_{meas} - u_{sm} 
\end{eqnarray}

Moreover the same approach allows to obtain the variance in the presence of spurious modulation
\begin{equation}\label{eq:q_var}
\begin{aligned}
\text{Var}[q] & = & \int_{0}^{2\pi}F(\varphi) \Big(2\cos(2\varphi)- E[q] \Big)^2 d\varphi \\
& = & \text{Var}[q_0] - \left( 2\mu p_0 \epsilon_{sm} \cos(2\varphi_{0})\cos(2\varphi_{sm}) + \epsilon_{sm}^{2} \cos^{2}(2\varphi_{sm}) \right)\\
& = & \text{Var}[q_0] - \text{Var}[q_{sp}] 
\end{aligned}
\end{equation}
\begin{equation}\label{eq:u_var}
\begin{aligned}
\text{Var}[u] & = & \int_{0}^{2\pi}F(\varphi) \Big(2\sin(2\varphi)- E[u] \Big)^2 d\varphi \\
& = & \text{Var}[u_0] -\left( 2\mu p_0 \epsilon_{sm} \sin(2\varphi_{0})\sin(2\varphi_{sm}) + \epsilon_{sm}^{2} \sin^{2}(2\varphi_{sm}) \right)\\
& = & \text{Var}[u_0] - \text{Var}[u_{sp}] 
\end{aligned}
\end{equation}
where the terms
\begin{eqnarray}
\text{Var}[q_{sp}]  & = 2\mu p_0 \epsilon_{sm} \cos(2\varphi_{0})\cos(2\varphi_{sm}) + \epsilon_{sm}^{2} \cos^{2}(2\varphi_{sm}) \\ 
\text{Var}[u_{sp}] &=2\mu p_0 \epsilon_{sm} \sin(2\varphi_{0})\sin(2\varphi_{sm}) + \epsilon_{sm}^{2} \sin^{2}(2\varphi_{sm}) 
\end{eqnarray}
take into account the variance due to the presence of spurious modulation. 

It is worth noting that in all practical situations the first term ($\text{Var}[q_0]$ 
and $\text{Var}[u_0]$ of Eq. \ref{eq:q0_var} and Eq. \ref{eq:u0_var}) is much 
larger than the second ($\text{Var}[q_{sp}]$ and $\text{Var}[u_{sp}]$), which 
can be neglected. 
In fact, even in the worst case scenario of a large source polarization ($\mu p_0\approx 0.2$) 
and large spurious modulation ($\epsilon_{sm} \approx 0.05$), which maximizes the latter 
contribution, the first term in Eq. \ref{eq:q_var} and Eq. \ref{eq:u_var} is $\approx$ 2, 
whereas the second is $\approx$ 0.006.

\section{Measurement of the response to unpolarized radiation\label{sec:unpol_response}}

\begin{figure}
\begin{centering}
\includegraphics[width=0.7\linewidth]{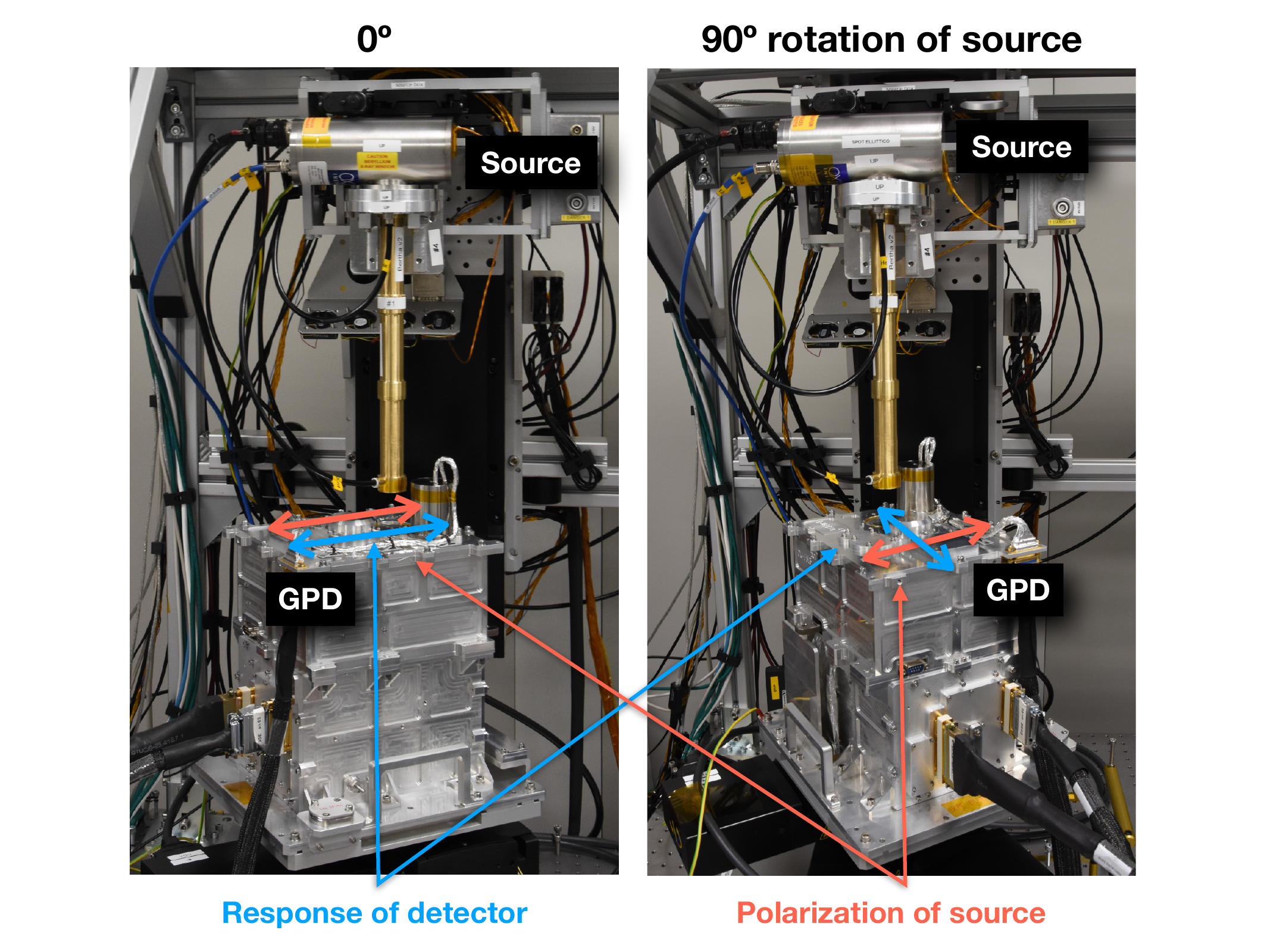}
\par\end{centering}
\caption{Rotation of the source to decouple the contribution to polarization
due to spurious modulation and due to the polarization of the source.
A rotation of 90$^{\circ}$, from the point of view of the detector, changes the sign of the Stokes parameters
of the source but not of spurious modulation. For this reason all
calibration measurements are taken at two orthogonal rotation angles.\label{fig:decoupling}}
\end{figure}

In this section we describe the method used to measure spurious modulation.
The laboratory sources used to do this measurement are partially polarized,
with the exact degree of polarization depending on a number of parameters 
which are difficult to estimate, such as the geometry of the emission internal 
to the source and the source spectrum (see \citet{MULERI2022102658}). 
Therefore, we define a procedure to decouple the intrinsic response of the
 instrument from the signal due to any genuine source polarization, 
which is based on repeating the measurement at two polarization angles shifted of 90$^\circ$. 
For these two measurements, from the GPD reference frame, the angle
of spurious modulation remains constant, while the angle of the intrinsic
modulation of the source changes by 90$^\circ$ (see Fig. \ref{fig:decoupling}).
Therefore the modulation caused by the true source polarization and
the spurious one will sum differently in the two measurements. 

Starting from the normalized Stokes parameters $q$ and $u$ (Eq.
\ref{eq:q} and \ref{eq:u}), for the two measurements we have

\begin{equation}
\begin{cases}
q_{0}=q_{sm}+q_{source}\left(\epsilon=0^{\circ}\right)\\
q_{90}=q_{sm}+q_{source}\left(\epsilon=90^{\circ}\right)=q_{sm}-q_{source}\left(\epsilon=0^{\circ}\right)
\end{cases}\label{eq:q0q90}
\end{equation}
\begin{equation}
\begin{cases}
u_{0}=u_{sm}+u_{source}\left(\epsilon=0^{\circ}\right)\\
u_{90}=u_{sm}+u_{source}\left(\epsilon=90^{\circ}\right)=u_{sm}-u_{source}\left(\epsilon=0^{\circ}\right)
\end{cases}\label{eq:u0u90}
\end{equation}
where $sm$ refers to the contribution of spurious modulation (independent
from the rotation angle), and $source$ refers to the contribution
due to the source (dependent on the rotation angle). The change of
sign of the source components after a $90^{\circ}$ rotation is due
to the fact that $q$ is defined as $q=2\cos\left(2\phi\right)$ and
analogously $u$ (Eq. \ref{eq:q_i} and \ref{eq:u_i}). Solving
the system of equations above gives
\begin{equation}
q_{sm}=\frac{q_{0}+q_{90}}{2}\label{eq:qsm}
\end{equation}

\begin{equation}
u_{sm}=\frac{u_{0}+u_{90}}{2}\label{eq:usm}
\end{equation}

with uncertainty given by
\begin{equation}
 \sigma_{q_{sm}} = \frac 1 2 \sqrt{\sigma_{q_{0}}^2 + \sigma_{q_{90}}^2}= 
 \frac{1}{2}\sqrt{\frac{2-q_{0}^2}{N_{0}-1}+\frac{2-q_{90}^2}{N_{90}-1}} \label{eq:qsm_err}
\end{equation} 
\begin{equation}
 \sigma_{u_{sm}} = \frac 1 2 \sqrt{\sigma_{u_{0}}^2 + \sigma_{u_{90}}^2}= 
 \frac{1}{2}\sqrt{\frac{2-u_{0}^2}{N_{0}-1}+\frac{2-u_{90}^2}{N_{90}-1}} \label{eq:usm_err}
\end{equation} 

Analogously it is possible to obtain the Stokes parameters of the source
$q_{source}$ and $u_{source}$ as a byproduct of this method

\begin{equation}
q_{source}=\frac{q_{0}-q_{90}}{2}
\end{equation}

\begin{equation}
u_{source}=\frac{u_{0}-u_{90}}{2}
\end{equation}

These equations will be used as a comparison to test the correction algorithm below
in Sec. \ref{sec:calibration_testing}.

The correction can be applied by subtracting from a measurement the spurious modulation 
calculated with Eq. \ref{eq:qsm} and Eq. \ref{eq:usm}).

This has however two disadvantages:
\begin{itemize}
\item GPD spurious modulation is position-dependent \citep{baldini2021}, 
and therefore photons used for calibration should be extracted from the same region 
as that for measurement. This is very unpractical.
\item If the photons selected have a non monochromatic spectrum, the correction
cannot easily take into account the energy dependence of spurious
modulation. 
\end{itemize}
These two problems can be overcome by correcting each event (i.e.
each photon) singularly. This method, which is the one that will
be used for celestial observations, is described in the next section.

\section{Description of the photon by photon algorithm\label{sec:ph_by_ph}}

\subsection{Creation of the calibration database\label{subsec:maps_creation}}

The creation of the calibration database is based on measurements illuminating 
the entire GPD sensitive area (flat field measurement), repeated at two source azimuthal angles 
one orthogonal with respect to the other and at different energies.

\begin{figure}
\begin{centering}
\includegraphics[width=0.5\linewidth]{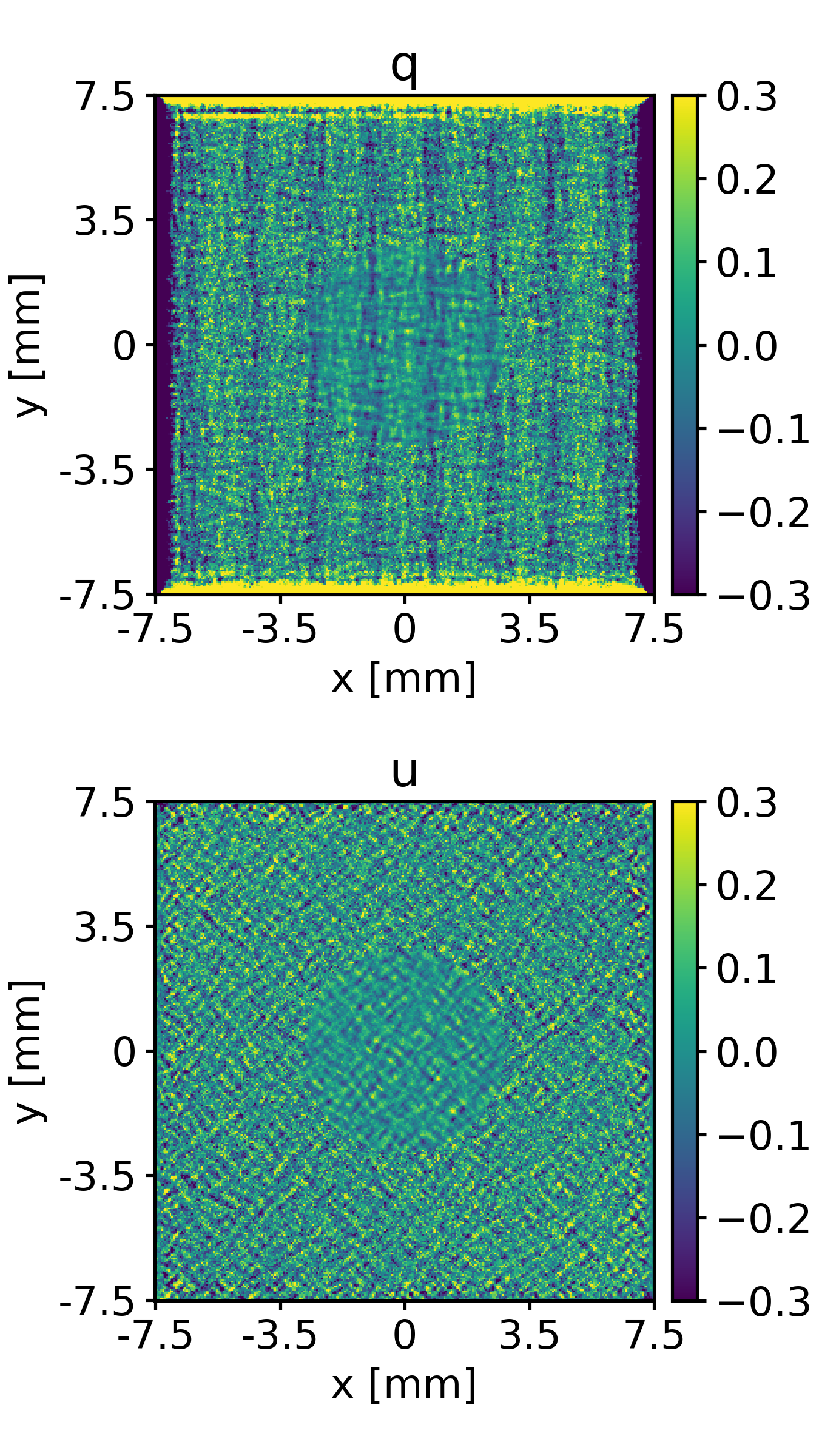}
\par\end{centering}
\caption{Example of spurious modulation maps for Stokes parameters $q$ and $u$
at 2.7~keV. The central region is smoother because it was calibrated with a larger statistics. 
It is also worth noting that values at the border are biased because photoelectric 
tracks are truncated by the physical size of the ASIC used for reading out the signal.
The bias is such that the angle is biased to 90$^\circ$ for vertical edges and to 0$^\circ$ for horizontal edges. 
$q$ is defined as the cosine of twice such angle (Eq. \ref{eq:q_i}) and, therefore, values are biased to
$\sim \cos(2\cdot90^\circ)$ and $\sim \cos(2\cdot0^\circ)$. $u$ (Eq. \ref{eq:u_i}), instead, is 
$\sim \sin(2\cdot90^\circ)$ and $\sim \sin(2\cdot0^\circ)$ and therefore the bias is around 0.
\label{fig:sm_maps}}
\end{figure}

For each energy and angle the events are divided, according to their
spatial positions (track estimated absorption points), in a certain number of bins (300$\times$300 in
this paper). In each spatial bin the normalized Stokes parameters are computed
from Eq. \ref{eq:q} and \ref{eq:u}. The Stokes parameters
for spurious modulation for each bin are then computed from Eq.
\ref{eq:qsm} and \ref{eq:usm}. These matrices are saved in a file,
which also contains the energy for each map. An example of these maps
is shown in Fig. \ref{fig:sm_maps}. 

The number of bins is chosen so to have sufficient granularity in
the source image. However, it should be noted that the correction
of the single event is by itself not significant, but the correction over many events becomes statistically functional.

\subsection{Spurious modulation removal}

A measurement is corrected photon by photon, subtracting from the Stokes
parameters of the event the values measured during calibration in
the spatial bin in which the photon is absorbed and at its measured energy.
To achieve this last point, Stokes maps at different energies are linearly interpolated bin by bin.

In practice, Eq. \ref{eq:q_i} and \ref{eq:u_i} are modified as

\begin{equation}
q\left[i\right]_{uncorrected}=2\cos\left(2\phi\left[i\right]\right)\label{eq:q_unc}
\end{equation}
\begin{equation}
u\left[i\right]_{uncorrected}=2\sin\left(2\phi\left[i\right]\right)\label{eq:u_unc}
\end{equation}
\begin{equation}
q\left[i\right]_{corrected}=q\left[i\right]_{uncorrected}-q_{sm}\left[bin_x\right]\left[bin_y\right]\left[energy\right]\label{eq:q_cal}
\end{equation}

\begin{equation}
u\left[i\right]_{corrected}=u\left[i\right]_{uncorrected}-u_{sm}\left[bin_x\right]\left[bin_y\right]\left[energy\right]\label{eq:u_cal}
\end{equation}
where $q_{sm}$ and $u_{sm}$ are the values of the maps created in
Sec. \ref{subsec:maps_creation}.

The Stokes parameters for all events under consideration are then
given by Eq. \ref{eq:q} and \ref{eq:u}, and modulation, polarization
and phase by Eq. \ref{eq:m}, \ref{eq:p} and \ref{eq:psi}.

\subsection{Uncertainty on calibrated modulation}\label{subsec:uncertainty}
\begin{figure}
\begin{centering}
\includegraphics[width=0.9\linewidth]{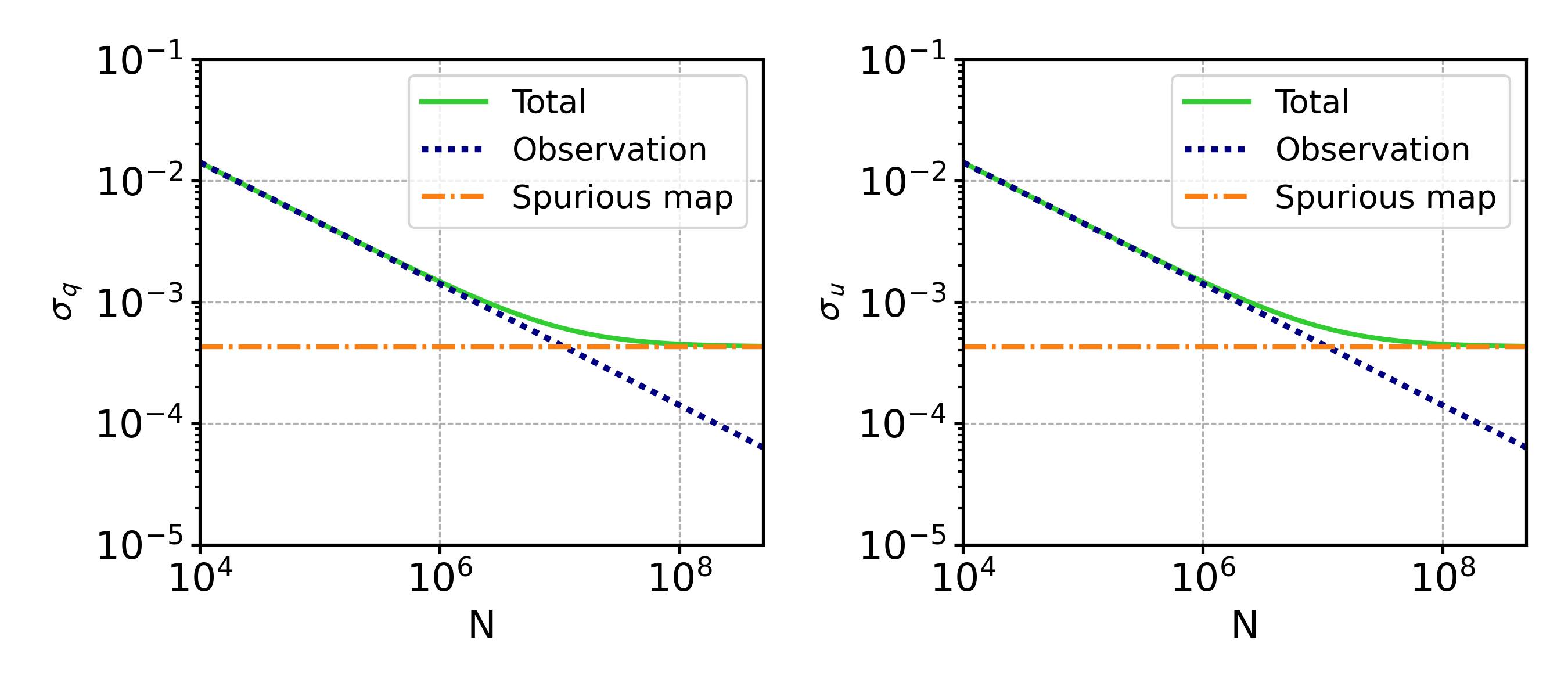}
\par\end{centering}
\caption{Comparison of the 1st term (Observation) and 2nd term (Spurious map) of the uncertainty expression (Eq. \ref{eq:IIterms}) 
as a function of the number of events of the observation. It can be seen that the 2nd term is often negligible. Also plotted  is the total quadratic sum of the two terms, that tends to an asintotic value which is the residual due only to calibration.
This example is at 2.7 keV with $7\cdot10^6$ events in the spurious modulation map.
 \label{fig:IIterms_err}}
\end{figure}

To understand how the uncertainties propagate it can be instructive to write the expression 
for the corrected normalized $q$ parameter including Eq. \ref{eq:q_cal} in Eq. \ref{eq:q} (the $u$ case is analogous)
\begin{equation}
q_{corrected}=\frac{\sum_{i}\left[q_{unc,i}\left(x_i, y_i, E_i\right)-q_{sm}\left(bin_x, bin_y, bin_E, E_i\right)\right]}{N_{obs}}
\end{equation}
where it is evident that each uncorrected event of the observation depends on the spatial position 
$x_i,y_i$ and on the measured energy $E_i$, while the subtracted spurious modulation depends on the bins 
(both the  spatial and energy calibration bins) and on the specific interpolated energy $E_i$ inside the bin. 
$N_{obs}$ is the number of events in the observation which should be distinguished from the number of
events in the calibration measurements.
 The expression above can be expanded by separating the sum over the spatial and energy bins $\sum_{bins}$ 
from the sum over the single events inside each bin $\sum_j$
\begin{equation}
q_{corrected}=\frac{\sum_{i}q_{unc,i}\left(x_i, y_i, E_i\right)}{N_{obs}}-\frac{\sum_{bins}\left[\sum_{j}q_{sm}\left(bin_x, bin_y, bin_E, E_j\right)\right]}{N_{obs}}
\end{equation}
It should be noted that $\sum_{bins}$ is a sum over uncorrelated terms, while $\sum_j$ is a sum 
over (partially) correlated terms (this is because the subtracted spurious modulation comes from the same bin). 
Looking at the uncertainty and (as a worst-case) assuming correlation $=1$ in the same spatial and energy bins,
$\sum_j$ disappears and one uncertainty is considered for each bin (this is a good approximation 
because the variation with energy due to interpolation in the same energy bin is small). The total uncertainty is
\begin{equation}
\sigma_{q_{corrected}}=\sqrt{\sigma_{q_{unc}}^{2}+\sum_{bins}\left(\frac{n_{obs}\left(bin_x, bin_y, bin_E\right)}{N_{obs}}\sigma_{q_{sm}}\left(bin_x, bin_y, bin_E\right)\right)^{2}} \label{eq:IIterms}
\end{equation}
where $\sigma_{q_{unc}}$ is given by Eq. \ref{eq:q_err},  $\sigma_{q_{sm}}$ is given by Eq. \ref{eq:qsm_err}, 
and $n_{obs}$ is the sum of the events of the observation in the spatial and energy bins.
The two terms are compared in Fig. \ref{fig:IIterms_err}, where it can be seen 
that the 2nd term will be negligible in most practical situations, 
as IXPE observations will typically have a number of counts smaller than calibrations. 
In this case, the simpler expressions of 
Eqs. \ref{eq:q_err}, \ref{eq:u_err}, \ref{eq:m_err} and \ref{eq:psi_err} can be 
used. The two terms are also further compared using simulations in Sec. \ref{subsec:simulations}.

The same plot also presents the quadratic sum of the two terms. It can be seen that the uncertainty value tends to an asintotic value, which is the residual uncertainty due only to calibration.

\begin{figure}
\begin{centering}
\includegraphics[width=0.8\linewidth]{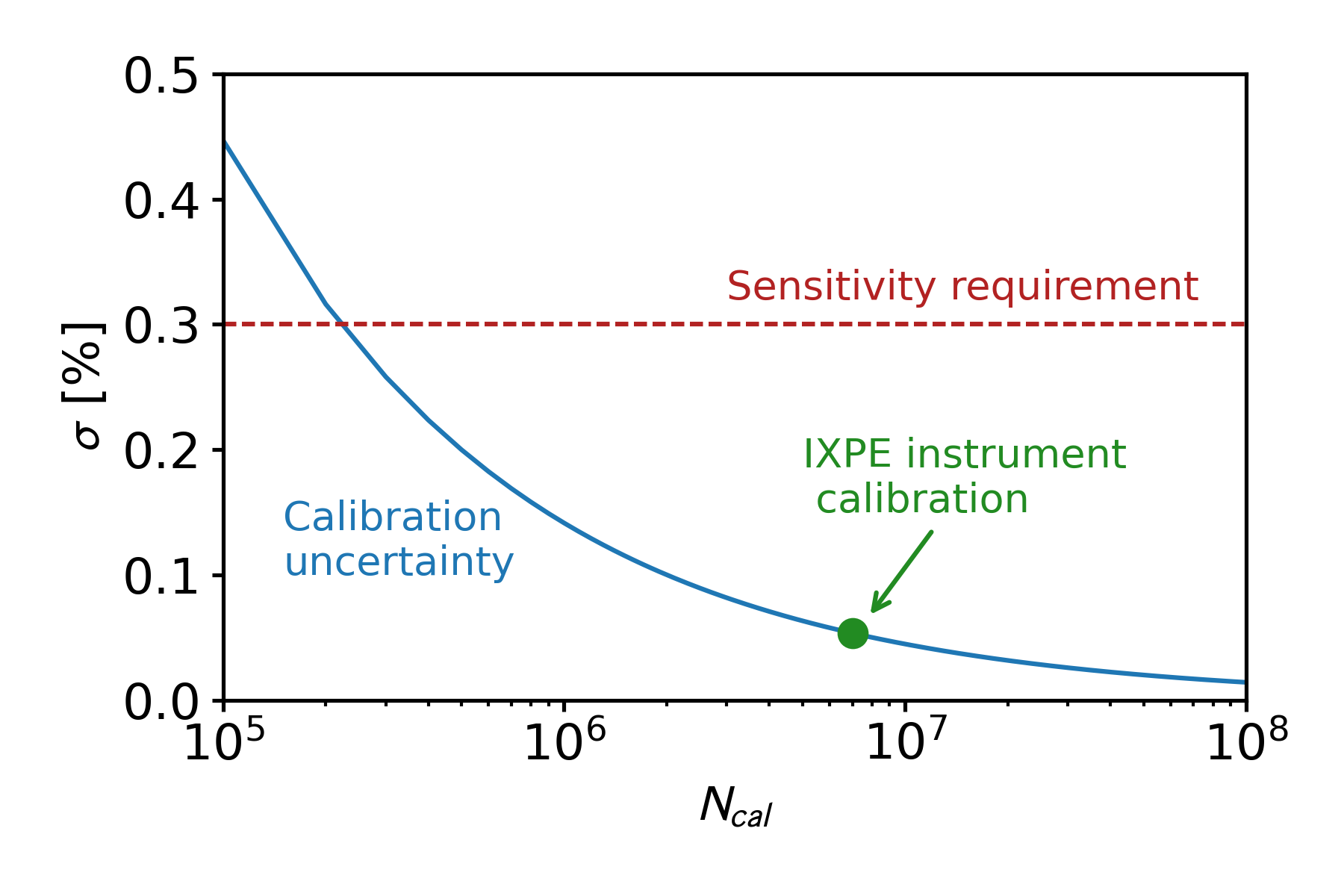}
\par\end{centering}
\caption{Uncertainty on modulation, as a function of the number of counts $N_{cal}$ in the calibration measurements, for  the ideal case of an infinite number of counts in the observation. The uncertainty in this case depends on the number of counts in the calibration measurement as $\sigma \approx \sqrt{ 2 / N_{cal}}$. Also shown in this plot are the IXPE requirement of 0.3\% modulation and the number of counts of IXPE's calibration measurement (at 2.7~keV but representative of the other energies). 
 \label{fig:cal_sensitivity}}
\end{figure}

To understand what this minimum uncertainty (proportional to the minimum detectable amplitude)
is, we can consider the ideal case of an infinite number of counts in the observation. In this ideal scenario the uncertainty on a modulation measurement depends only on the number of counts $N_{cal}$ in the calibration measurements. For IXPE's GPDs, in which spurious modulation is not too high, Eq. \ref{eq:m_err} can be approximated as $\sigma \approx \sqrt{ 2 / N_{cal}}$. This last expression also gives the same result as the II term of Eq. \ref{eq:IIterms} if the  observation is spatially uniform and monochromatic. This uncertainty is plotted in Fig. \ref{fig:cal_sensitivity}, where it is shown that  IXPE's instrument calibration has been made with enough counts to be well inside the requirement. The actual sensitivity will be worse than this estimate because of small temporal variations, but still well inside the requirement \citep{spurious_future}.

\section{Testing the spurious modulation correction\label{sec:calibration_testing}}

\subsection{Modulation of monochromatic laboratory sources}\label{sec:decoupling_comp}

To test the effectiveness of the correction we compare some results of on-ground calibration 
obtained with the photon-by-photon algorithm described in Sec. \ref{sec:ph_by_ph}, with 
the results of the "standard" analysis described in Sec. \ref{sec:unpol_response}, here named global decoupling. 

\begin{figure}
\begin{centering}
\includegraphics[width=0.8\linewidth]{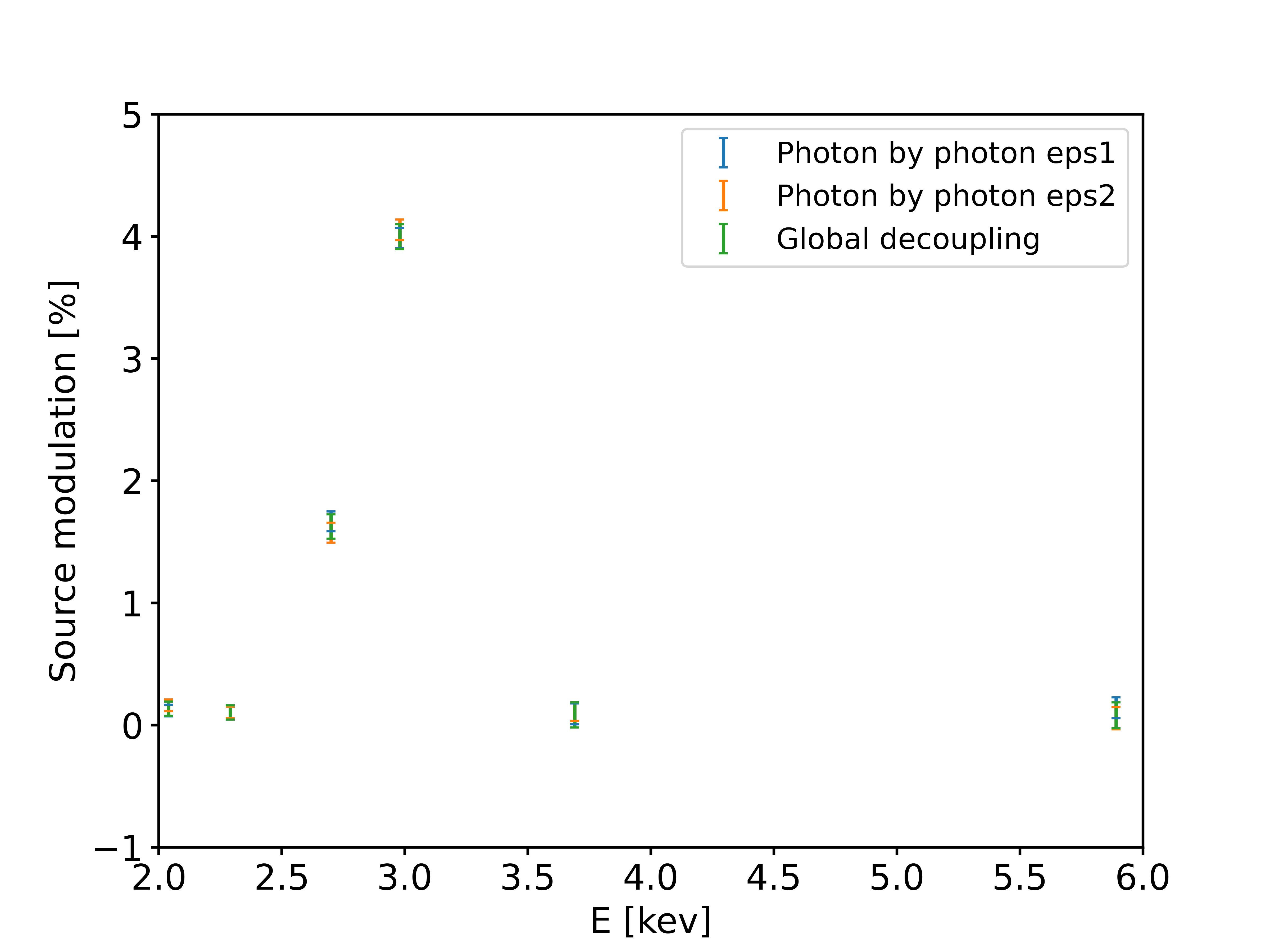}
\par\end{centering}
\caption{Modulation of low polarization flat field laboratory sources at various
energies. One value is referred to the global decoupling, the other
two values are the application of the photon by photon algorithm to
the flat fields at the two rotation angles (named \emph{eps1} and \emph{eps2}) at which this measurement 
was acquired (hence the two values). All three
values are compatible.\label{fig:UnpFF_comparison}}
\end{figure}

We show in Fig. \ref{fig:UnpFF_comparison} the modulation due to the genuine polarization 
of the sources used for the calibration of IXPE DUs, derived as by-product of the calibration 
of the detector response to unpolarized radiation. 
One value is referred to the global decoupling, the other two values
are the application of the photon by photon algorithm to the two measurements 
rotated of 90 degrees. (hence the two values). All three values are compatible.

\begin{figure}
\begin{centering}
\includegraphics[width=0.7\linewidth]{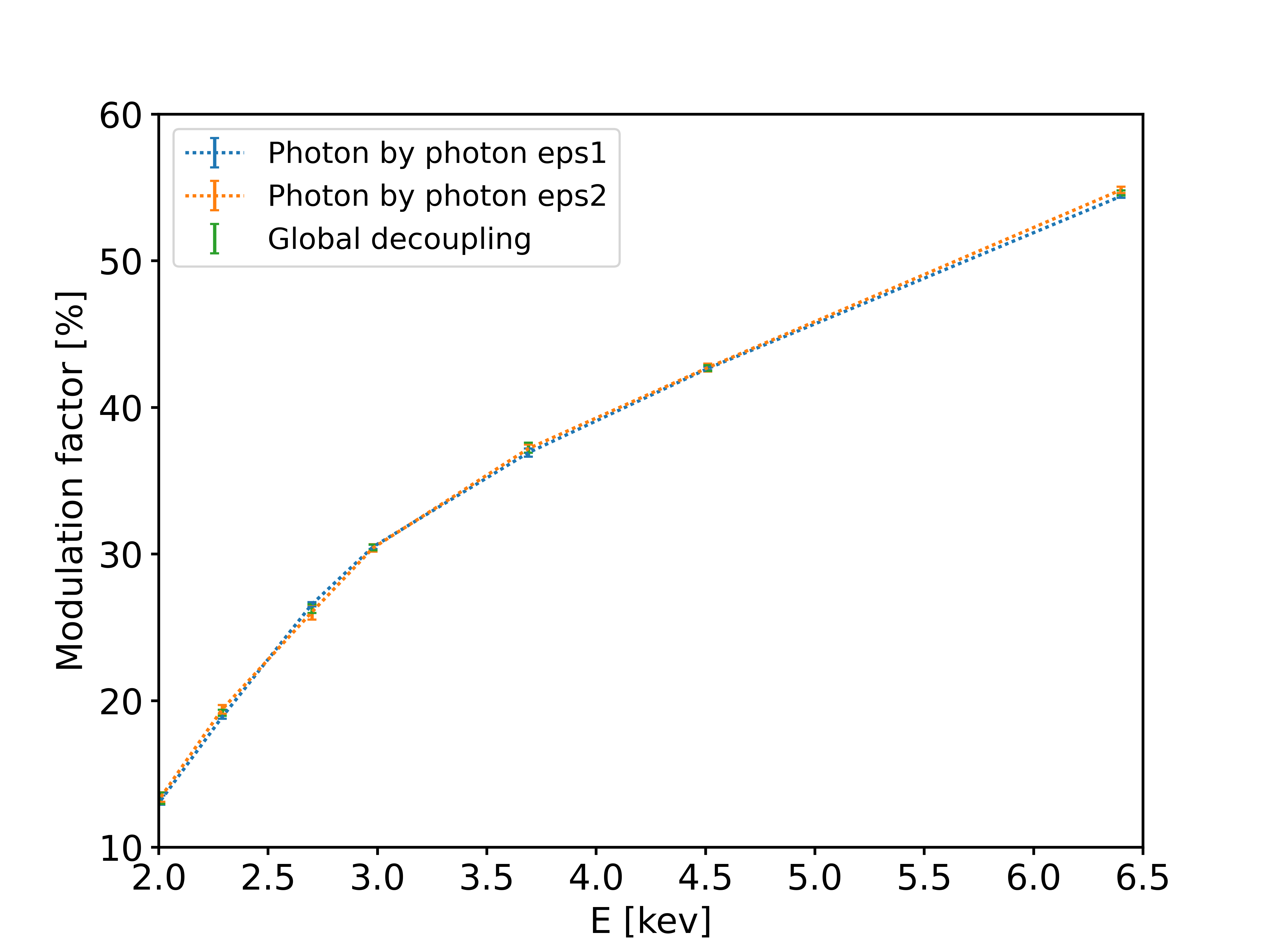}
\par\end{centering}
\caption{Same as Fig. \ref{fig:UnpFF_comparison} but for the modulation factor 
(obtained by dividing the source modulation of the polarized measurements by the source polarization, as in Eq. \ref{eq:p}).\label{fig:Pol_comparison}}
\end{figure}

In Fig. \ref{fig:Pol_comparison} we compare the value of modulation factor, 
measured with polarized sources, 
obtained with the two different methods.
All three values are again compatible.

These tests prove that, for monochromatic sources, the photon by photon
correction method retrieves the correct modulation.

\subsection{Spectrally resolved polarization}

One of the sources used for ground calibration is the Ca X-ray tube,
whose spectrum comprises of calcium fluorescence lines plus a significant contribution 
from continuum bremsstrahlung emission. This source is expected to be completely 
unpolarized from first principles (see \citet{MULERI2022102658}), and also 
gives the opportunity of testing the algorithm for spurious modulation removal with 
non-monochromatic sources. In this case it is
necessary to correct for spurious modulation each single photon with
its specific measured energy, as is done by the photon by photon algorithm. The Stokes
parameters and subsequent modulation can then be obtained for photons
in different energy bins. 

\begin{figure}
\begin{centering}
\includegraphics[width=0.7\linewidth]{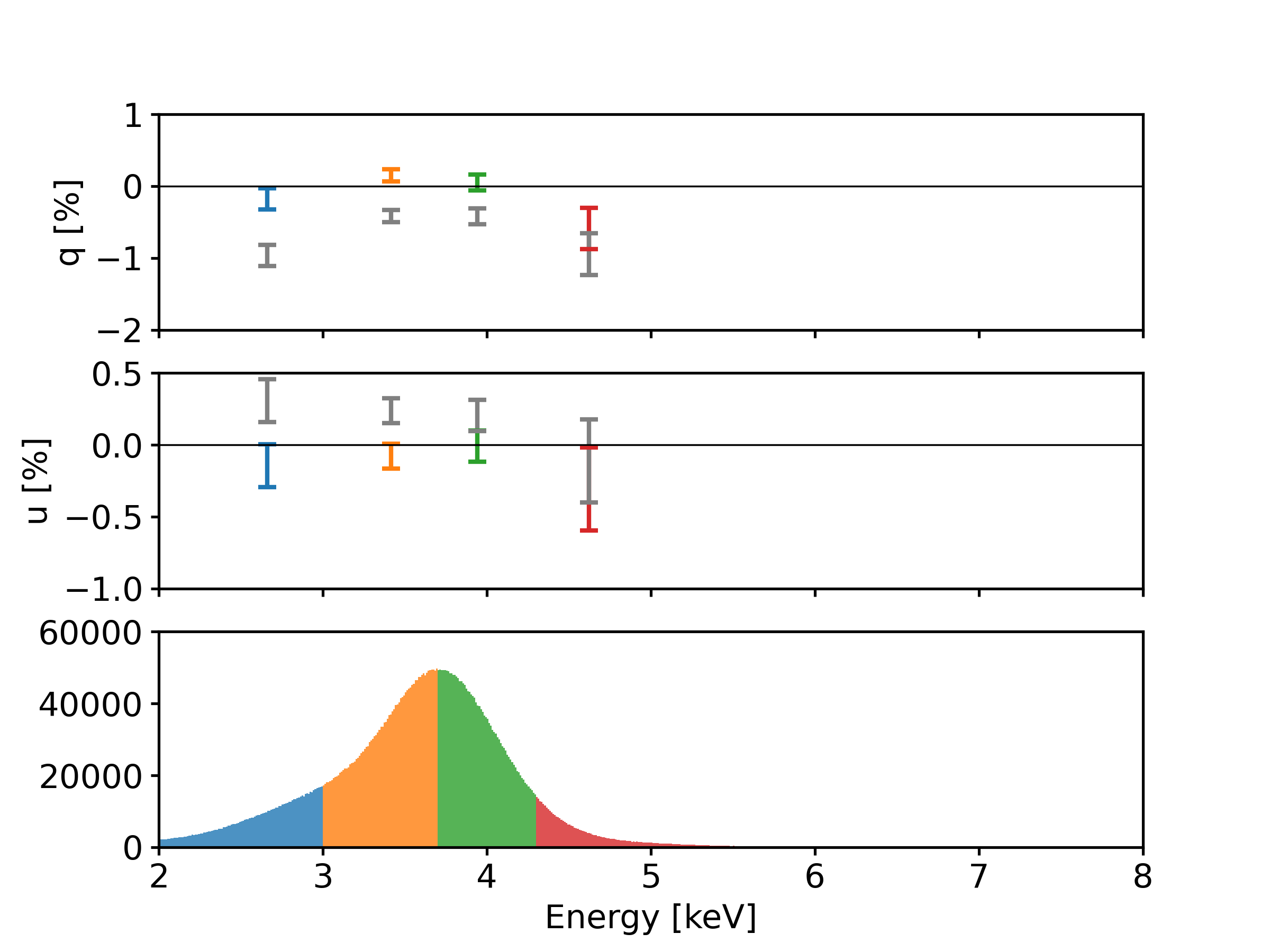}
\par\end{centering}
\caption{Spectra of the Ca X-ray tube with the modulation computed in energy
bins. The colored points are corrected for spurious modulation using
the photon by photon correction (therefore sensitive to the different
energies), while the gray points are uncorrected. The corrected
points are compatible with 0 as expected for the Ca X-ray tube, while
for the uncorrected points the decreasing trend of spurious modulation
with energy is evident. \label{fig:Ca_spectra}}
\end{figure}

This is shown in Fig. \ref{fig:Ca_spectra}. The colored data points
are corrected for spurious modulation, and are compatible with 0 as
expected, while the gray points are the uncorrected points.

Two points should be noted from this result. First, this is a situation
in which only the photon by photon algorithm can be used, which therefore
demonstrates its more general applicability. Second, the uncorrected
points show a slightly decreasing trend (consequence of the fact that spurious
modulation decreases with energy, see \citet{spurious_future}); this
trend disappears after correction, proving that the algorithm is removing
the systematic effect correctly.

\subsection{Effect of finite energy resolution}

The GPD, as any other real detector, measures the energy of the event with a 
finite energy resolution. Spurious modulation changes with the true energy 
of the radiation, but is corrected starting from the \emph{measured} energy. 
In this section we evaluate the effect, if any, for a representative observation 
simulated with the detailed GPD Monte Carlo software developed by the IXPE team \citep{montecarlo}.

The simulation proceeds as follows. The first step is to input a Crab-like 
spectrum (a power law with index 2) in the Monte Carlo to produce 
photoelectron track images equivalent to those obtained with real 
measurements with the GPD. These are reprocessed with the same 
software used for real data. Simulations are able to reproduce all the 
characteristics of a real measurement in fine details, with the exception 
of spurious modulation. Such a component is added event-by-event, 
interpolating the maps of spurious modulation (obtained with calibration) 
at the true photon energy provided by the Monte Carlo. In the following 
step, spurious modulation is subtracted but its value is obtained by 
interpolating the calibration maps at the measured energy.  

\begin{figure}
\begin{centering}
\includegraphics[width=0.8\linewidth]{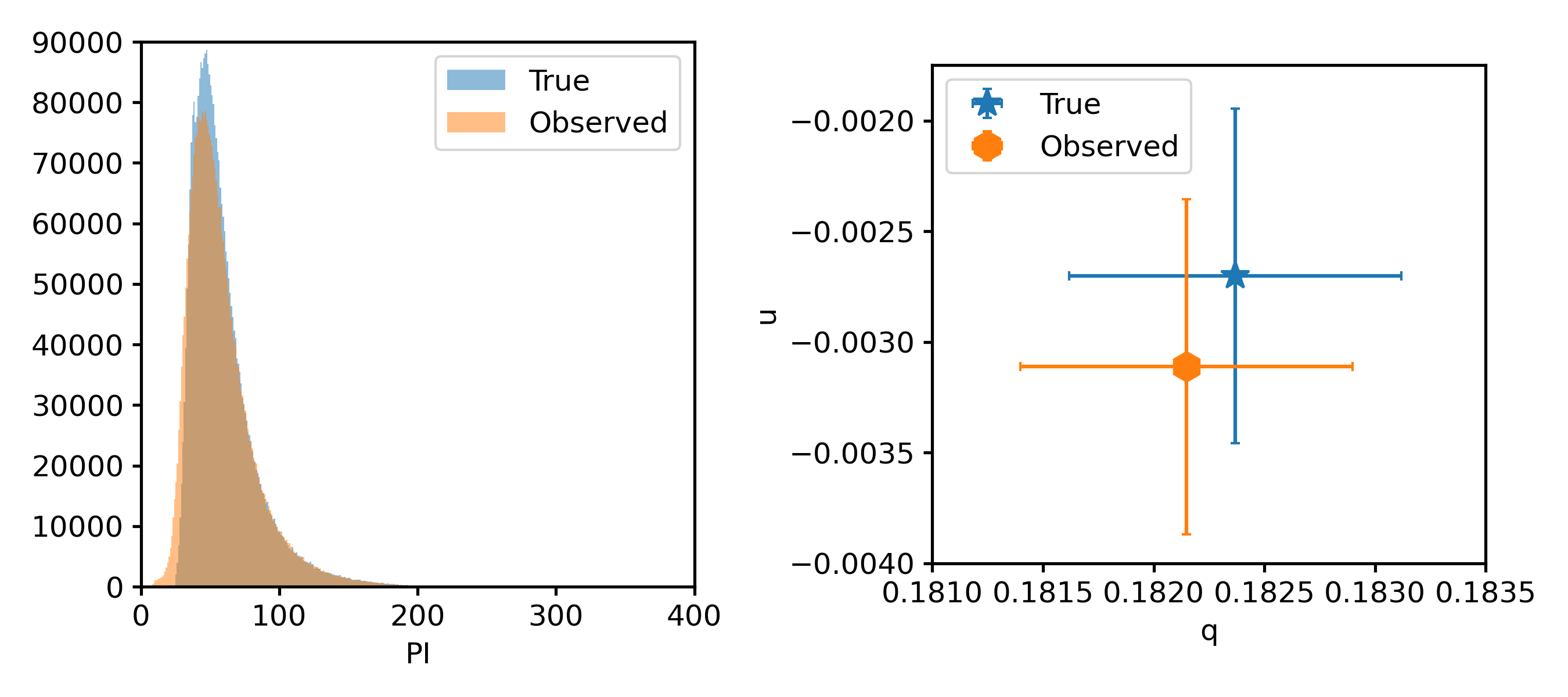}
\par\end{centering}
\caption{Comparison of modulation corrected using the true 
and observed energy using a 3.5M event Monte Carlo
simulation of a Crab-like spectrum. The plot on the left 
compares the spectra, and the one on the right
the Stokes parameters of the observation.
Spurious modulation is first summed by interpolating the calibration maps using the true energy
(given in input to the Monte Carlo), and then 
subtracted by interpolating using the measured energy. 
As can be seen, the offset in the correction due to the discrepancy 
between the two energy estimates is negligible.
 \label{fig:crab_spectra_comparison}}
\end{figure}

The measured modulation is compared with the expected value in 
Fig. \ref{fig:crab_spectra_comparison}. The latter is derived by 
processing the data obtained from the Monte Carlo without adding 
(and removing) spurious modulation. It is evident that the two values 
agree very well. The comparison of the source spectrum in true and 
measured energy is shown in the same figure.

\subsection{Spatial uniformity of the correction}
\begin{figure}
\begin{centering}
\includegraphics[width=0.7\linewidth]{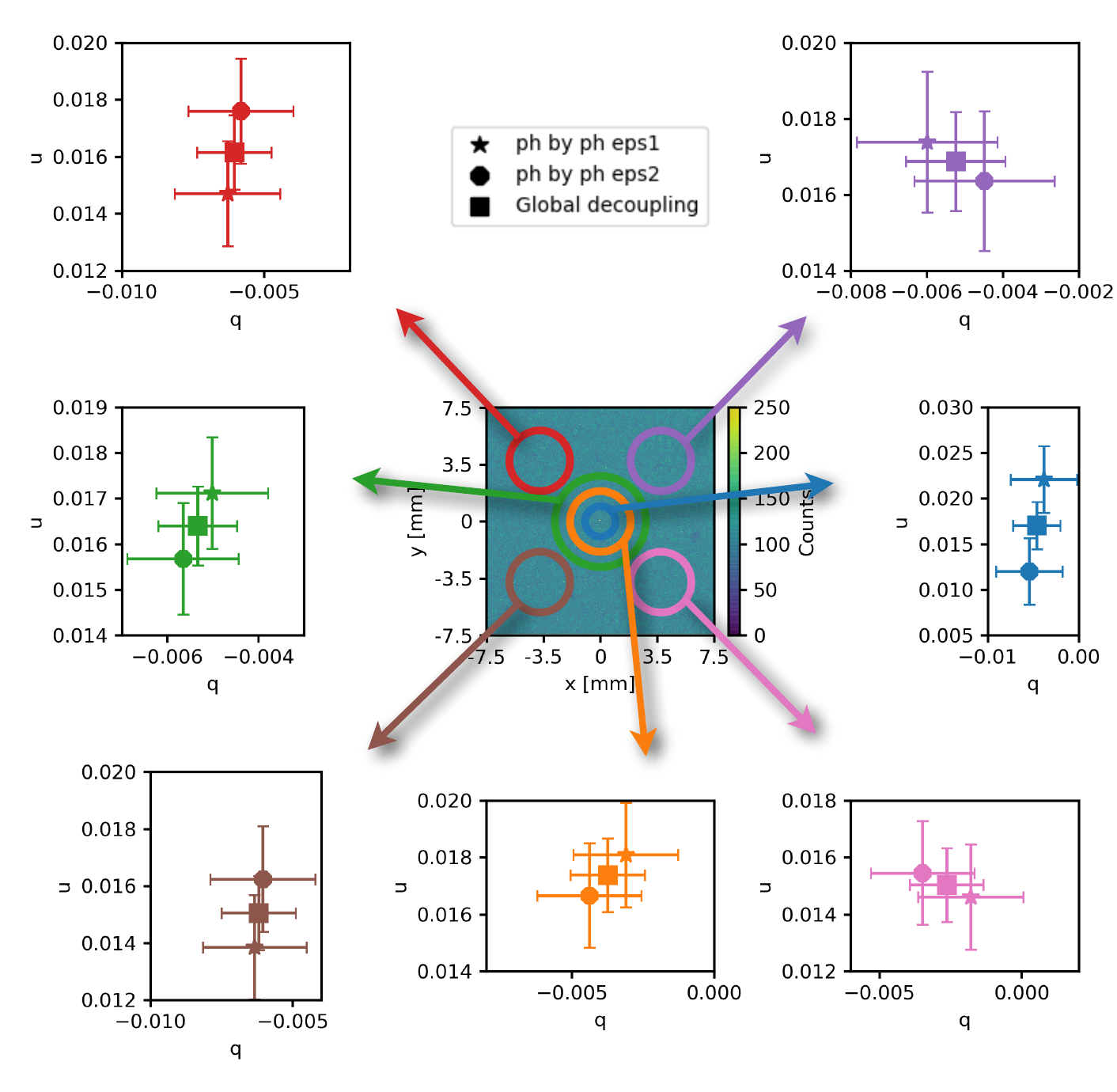}
\par\end{centering}
\caption{Modulation generated by the genuine source polarization at 2.7~keV, 
obtained by correcting spurious modulation with different methods in several regions of the detector. 
One value is referred to the global decoupling, the other
two values are the application of the photon by photon algorithm to
the flat fields at the two rotation angles (named \emph{eps1} and \emph{eps2}) 
at which this measurement 
was acquired (hence the two values). All three
values are compatible over all the detector, showing that the correction algorithm 
performs the same in different regions of the detector and/or with different extraction radii.
 \label{fig:spatial_trend}}
\end{figure}

The GPD is a detector with imaging capabilities, and therefore the 
correction algorithm will be applied in different parts of the detector, 
and also in regions of different sizes. To verify the spatial uniformity of 
the correction, the comparison between the correction algorithm and 
the standard analysis (global decoupling), as done in Sec. \ref{sec:decoupling_comp}, 
was repeated, for the measurement at 2.7~keV, in spatial regions of 
different position and size. As can be seen in Fig. \ref{fig:spatial_trend}, 
the algorithm performs well the corrections over the entire detector.

\subsection{Simulations\label{subsec:simulations}}

To further test the statistical distribution of the calibrated measurements, 
we simulated the application of spurious modulation calibrations to toy observations. 
Both the calibration measurements and monochromatic observations (see precise definitions below)
were simulated, mimicking the process that will be used in reality. To each observation a 
spurious modulation (dependent only on energy) is added (and then removed 
using the simulated calibrations). The goal of 
these simulations is to prove that, by subtracting the calibration 
measurements from the observations, the true modulation value 
is retrieved with the correct uncertainty as calculated in Eq. \ref{eq:IIterms}.

\subsubsection{Simulation procedure}

We define two types of simulated measurements:

\begin{enumerate}
\item ``Calibration'': simulated calibration measurement analogous to the ground 
measurements carried out to calibrate the response of the IXPE detectors. 
In the following we focus the discussion only in a subrange of the energy range 
which was effectively calibrated, that is, we simulated just two measurements at 2.7 and 2.98~keV. 

Analogously to real measurements, we simulated two measurements at 
orthogonal angles, each consisting of $15 \times 10^6$~events  at 2.7 and 2.98~keV 
and characterized by a polarization equal to the measured one. From these measurements, 
spurious modulation is derived as described in Sec \ref {sec:unpol_response}, with an 
uncertainty given by Eqs. \ref{eq:qsm_err} and \ref{eq:usm_err} (see details below).

\item ``Observation'': simulated IXPE observation of a (toy) celestial source. 
In the following we assume that the source is monochromatic, with energy 
between 2.7 and 2.98~keV, and that its true polarization is $q=0.04$ and $u=0.02$, 
which is representative of the weakly-polarized sources which will be observed by IXPE.
\end{enumerate} 

A simulated measurement, either for calibration or toy observation, 
is generated assuming a distribution 
of the photoelectric position angles obtained from a 
$\cos^2$ distribution with the desired modulation and phase, and 
an energy obtained from a Gaussian distribution with width given 
by the approximate real energy resolution of the detector

Each simulated measurement is a list of events, 
each consisting of its Stokes parameters and energy. To the Stokes
parameters a toy spurious modulation is summed, derived by 
phenomenologically fitting the real dependence as the inverse of energy. 

The observations (second point above) are then corrected, using the calibration measurements 
(first point), by applying the photon by photon subtraction method described in this paper.

\subsubsection{Distribution of observations with one calibration dataset}

We first investigated the case of the statistical distribution of the polarization obtained 
by a corrected observation at energy $E=2.8$~keV in case only one calibration 
dataset is available, which is the real case. $10^{4}$ observations with $10^{7}$ events 
each were simulated and corrected (using the same calibration dataset for all). In this case the uncertainty on calibration 
measurements (2nd term of Eq. \ref{eq:IIterms}) is comparable with the statistical 
uncertainty on observations (1st term of Eq. \ref{eq:IIterms}), but this will not be the case in most practical situations. 

\begin{figure}
\begin{centering}
\includegraphics[width=0.7\linewidth]{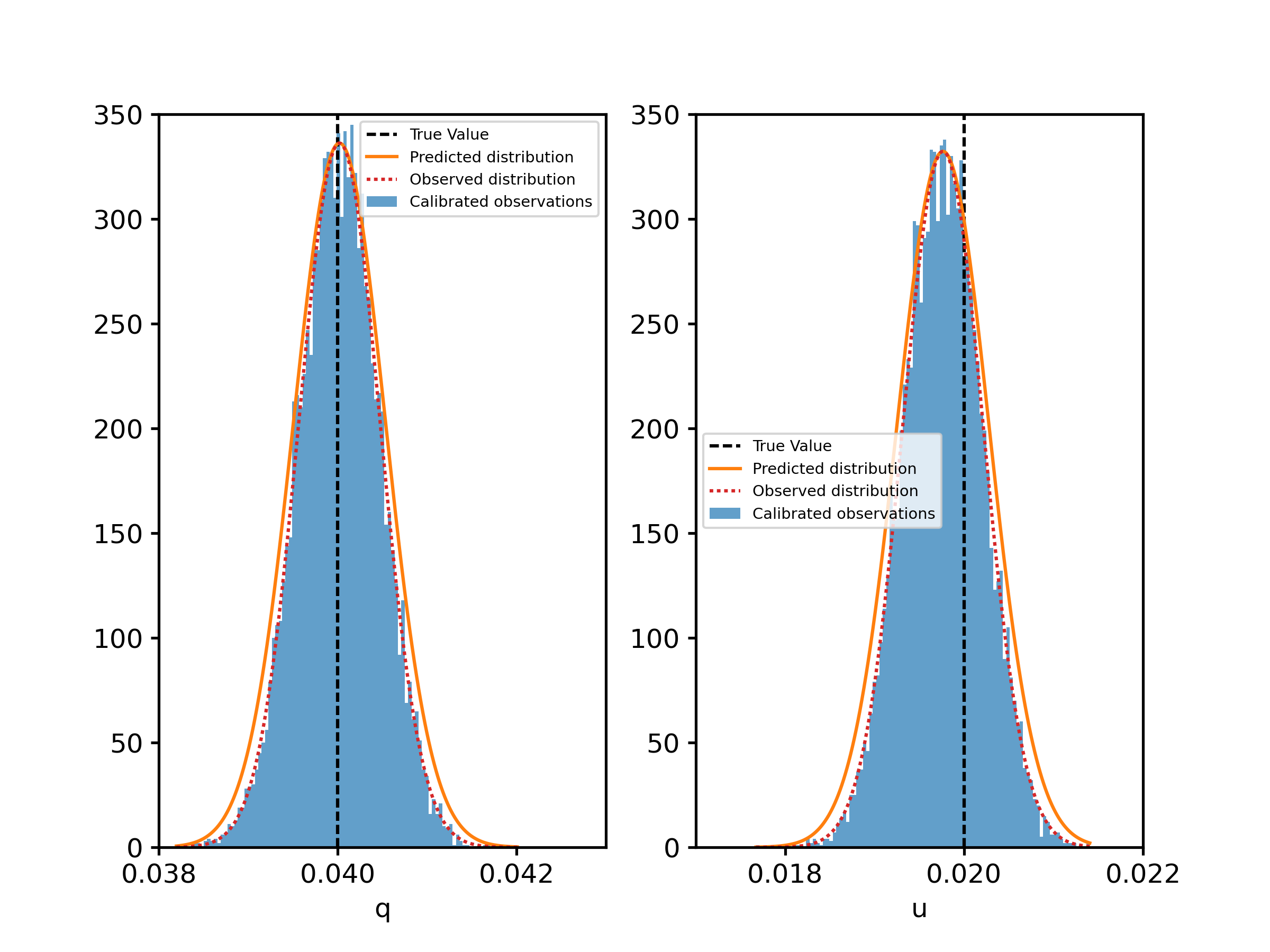}
\par\end{centering}
\caption{Observation at 2.8~keV consisting of $10^{4}$ iterations with $10^{7}$
events in each iteration. The predicted distribution of the observed values (orange line) 
is compared with the fit to the observed distribution (red line). The former has a larger 
width, because it accounts for the fact that in correcting we are subtracting an estimate 
of spurious modulation and not its true value. This causes a shift between 
the true modulation (vertical line) and the center of the observed distribution. \label{fig:sim_1_calibration}}
\end{figure}

The distribution of the derived source polarization is shown in Fig. \ref{fig:sim_1_calibration}. 
While values are Gaussian-distributed with a width derived from the number of counts in 
the observation as expected, the average value is shifted (this is particularly evident for $u$). 
This shift is due to the fact that we subtracted an \emph{estimate} of the spurious modulation 
amplitude and not its \emph{true} value. As we will see in the following, the 2nd term in 
Eq. \ref{eq:IIterms} accounts for this quite well.

In most IXPE observations the number of events will be much smaller than
the number of calibration events. In this case the uncertainty on
calibration measurements is negligible compared with the statistical
uncertainty on the observations, and so the difference between the estimated and true value 
of subtracted spurious modulation will also be negligible. In this assumption, the observed and 
predicted distributions would be essentially coincident.

\subsubsection{Distribution of observations with many calibration datasets}

In the previous section we have seen that corrected observations have a (small) offset with 
respect to the true value. We associated such a difference to the fact that we can remove 
from the observation just an estimate of the spurious modulation derived from calibration, 
and not its true value. To substantiate this claim, we investigate the observed values when 
the observations are corrected with many independent calibrations, which are then 
statistically distributed around the true value.

To study this $10^{4}$ observations with 
$10^{6}$ events each were again simulated (numbers chosen to keep the simulation 
running time reasonable). This time, however, each of these 
was corrected using $10^3$ different calibration measurements, that is, it was corrected $10^3$ times.
All of this was repeated at 2.7~keV and 2.98~keV (the calibration energies) and at other 
intermediate energies in the chosen 2.7-2.98~keV energy range.

\begin{figure}
\begin{centering}
\includegraphics[width=0.7\linewidth]{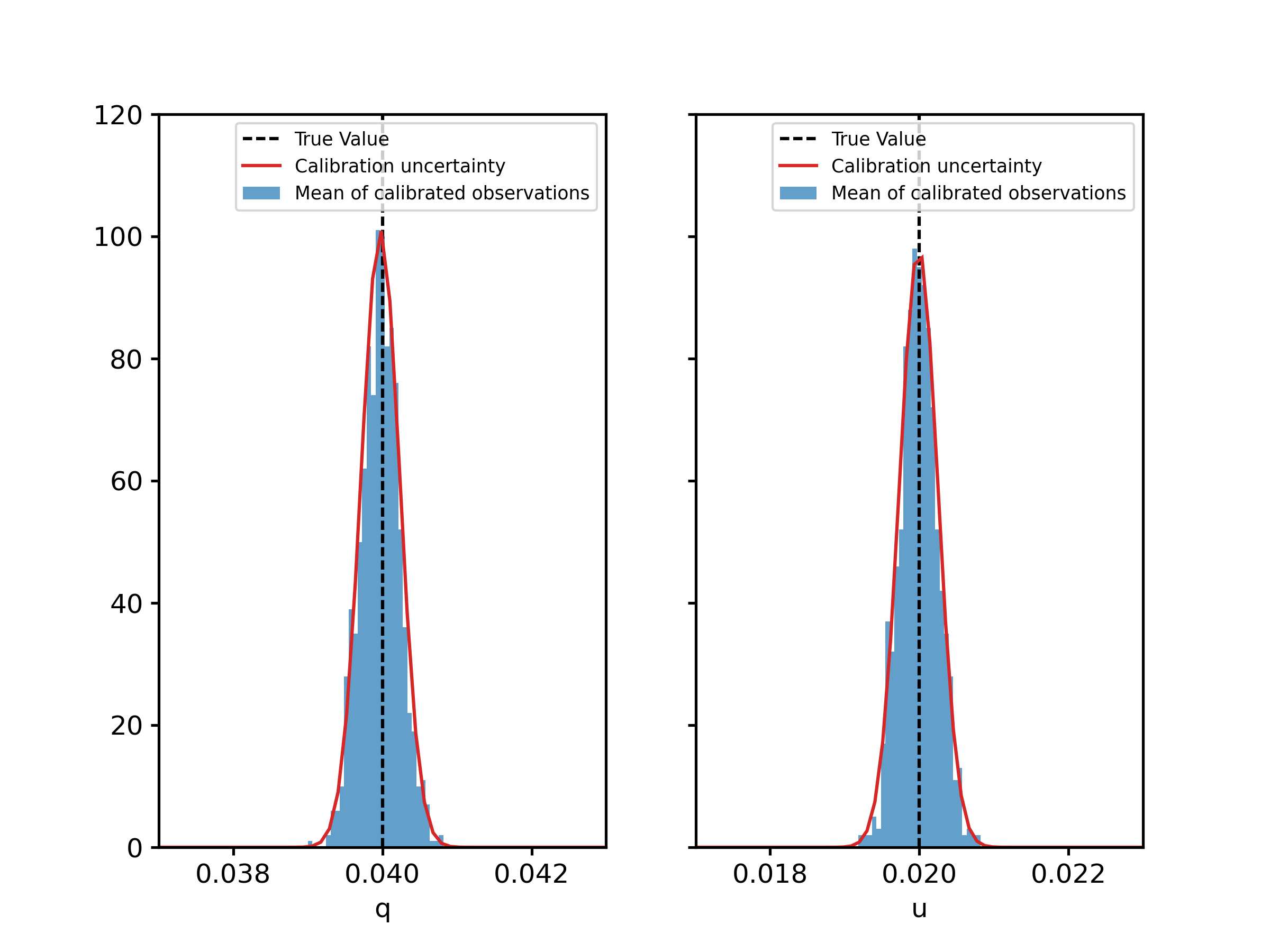}
\par\end{centering}
\caption{Distribution at 2.7~keV of the centers of observations (each consisting
of $10^{4}$ iterations of which each with $10^{7}$ events) corrected
with $10^{3}$ different calibration measurements. The difference with the distribution
of Fig. \ref{fig:sim_1_calibration} is that this is a distribution of the centers of $10^3$
distributions such as that of Fig. \ref{fig:sim_1_calibration}.
The distribution is correctly centered on the true value, with a width given by the calibration 
uncertainty (2nd term of Eq. \ref{eq:IIterms}).
\label{fig:sim_multiple_calibs_2.7}}
\end{figure}

Each of the $10^{3}$ sets of calibrated observations was fitted to derive 
the center of the observed source polarization  for the different calibrations. The distribution 
of such values is shown in Fig. \ref{fig:sim_multiple_calibs_2.7}. 

\begin{figure}
\begin{centering}
\includegraphics[width=0.7\linewidth]{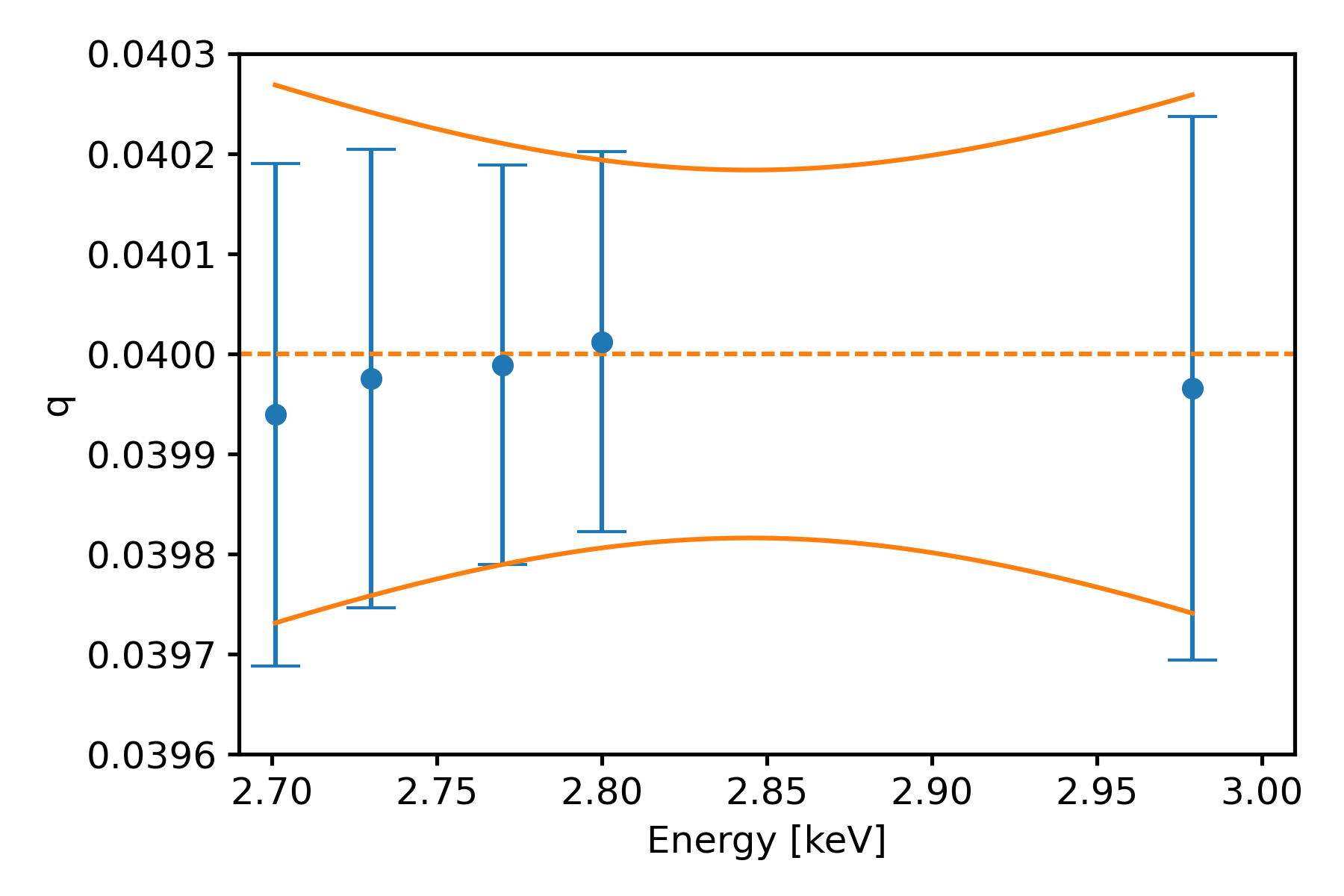}
\par\end{centering}
\caption{Centers and widths of Stokes parameter $q$ (the $u$ case is very similar) 
obtained from fitting the distributions of calibrations, such as the one in 
Fig. \ref{fig:sim_multiple_calibs_2.7}, at different energies. The orange lines 
are the widths computed with Eq. \ref{eq:sigma_vs_E}, centered on the true 
value (dashed line). The reason for the decrease in width is that the values 
in-between calibration energies are linearly interpolated, and the weighted 
combination of Eq. \ref{eq:sigma_vs_E} is more likely to be close to the center 
if far away from the calibration energies.
\label{fig:uncertainties_vs_energy}}
\end{figure}

The distribution is centered on the true modulation of
the source. This was expected because, averaging many independent calibrations, 
the value of spurious modulation which is corrected will tend to the true value. 
However the width of the distribution varies. The fitted standard deviation of the 
distributions is 0.026~\% at 2.7~keV,  0.023\% at 2.73~keV,
0.021\% at 2.77~keV, 0.018\% at 2.8~keV and 0.026~\% at 2.98~keV (see Fig. 
\ref{fig:uncertainties_vs_energy}). The amplitude of the uncertainty on calibration
(2nd term of equation \ref{eq:IIterms}) is 0.026\%, which is an upper limit to the 
values found at intermediate energies (this upper limit was derived in the 
worst-case assumption of perfectly correlated contributions in the same spatial and spectral bins).

The fact that the modulation values at 2.7~keV and 2.98~keV coincide with the 
calibration uncertainty (values written above) proves that the 2nd term
of equation \ref{eq:IIterms} is correct in estimating the uncertainty
at the calibration energies.

The discrepancy at the other energies is due to the fact that spurious modulation 
is linearly interpolated between the values measured at the energies of calibration 
(2.7 and 2.98~keV), and loses the assumed perfect correlation of the spatial and 
energy bins in Eq. \ref{eq:IIterms}.
The statistical uncertainty 
on the interpolated value can be obtained with a weighted combination of the 
values measured at the boundaries:
\begin{equation}
\sigma (E) = \sqrt{\left(\frac{2.98\text{keV}-E}{2.98\text{keV}-2.7\text{keV}}\sigma_{cal}^{2.7~\mathrm{keV}}\right)^{2}+\left(\frac{2.7\text{keV}-E}{2.98\text{keV}-2.7\text{keV}}\sigma_{cal}^{2.98~\mathrm{keV}}\right)^{2}}
\label{eq:sigma_vs_E}
\end{equation}

This expression correctly reproduces the observed values (see Fig. \ref{fig:uncertainties_vs_energy}).

\section{Conclusion\label{sec:conclusion}}

In this paper we presented the procedure to calibrate and correct the response 
to unpolarized radiation of the GPD.
First, the method to measure the response of the detector
to unpolarized radiation, using two measurements of the same source
rotated orthogonally, was presented. Then we discussed a correction algorithm
which corrects the systematics for each single event;
this allows great flexibility in the subsequent analysis.

The correction done with the photon by photon algorithm for monochromatic
sources was compared to that obtained as a byproduct of the unpolarized
response measurement; the two were shown to provide statistically
compatible results. The photon by photon algorithm was then analyzed
further using calibration data and simulations, proving its spectral 
capabilities and showing how the correction removes the trend of the 
systematic effect with energy. This demonstrates that 
the algorithm is able to subtract the systematic effect, achieving
all the sensitivity possible with the Gas Pixel Detector.

\begin{acknowledgments}
The Italian contribution to the IXPE mission is supported by the Italian Space Agency 
(ASI) through the contract ASI-OHBI-2017-12-I.0, the agreements ASI-INAF-2017-12-H0 
and ASI-INFN-2017.13-H0, and its Space Science Data Center (SSDC), and by the 
Istituto Nazionale di Astrofisica (INAF) and the Istituto Nazionale di Fisica Nucleare (INFN) in Italy. 

The United States contribution to the IXPE mission is supported by NASA as 
part of the Small Explorers Program. 
\end{acknowledgments}

\vspace{5mm}
\facilities{Imaging X-ray Polarimetry Explorer (IXPE)}

\bibliography{References}{}
\bibliographystyle{aasjournal}

\end{document}